\documentclass[superscriptaddress,aps,prl,twocolumn,floatfix]{revtex4-2}

\usepackage{graphicx}
\usepackage{bm}
\usepackage{dsfont}
\usepackage{amsmath, amsthm, amssymb,mathrsfs}
\usepackage{setspace}
\usepackage[utf8]{inputenc}
\usepackage[toc,page]{appendix}
\usepackage{epstopdf}
\usepackage{hyperref}
\usepackage{mdframed}
\usepackage{soul,xcolor}
\sethlcolor{yellow}

\newcommand{\ket}[1]{\left| #1 \right\rangle}

\newcommand{\ketbra}[2]{\left|#1\right\rangle\hskip-1mm\left\langle #2\right|}

\begin{document}

\title{Comment on ``Scheme of the arrangement for attack on the protocol BB84"}
\author{Jonte Hance}
\email{jonte.hance@bristol.ac.uk}
\affiliation{Quantum Engineering Technology Laboratory, Department of Electrical and Electronic Engineering, University of Bristol, Woodland Road, Bristol, BS8 1UB, UK}
\author{John Rarity}
\affiliation{Quantum Engineering Technology Laboratory, Department of Electrical and Electronic Engineering, University of Bristol, Woodland Road, Bristol, BS8 1UB, UK}

\begin{abstract}
    In a recent paper (\textit{Scheme of the arrangement for attack on the protocol BB84}, Optik 127(18):7083-7087, Sept 2016), a protocol was proposed for using weak measurement to attack BB84. This claimed the four basis states typically used could be perfectly discriminated, and so an interceptor could obtain all information carried. We show this attack fails when considered using standard quantum mechanics, as expected - such ``single-shot" quantum state discrimination is impossible, even using weak measurement.
\end{abstract}

\maketitle

In his recent paper, Khoklhov claims to have developed a protocol that can be used to distinguish between the four basis states used in BB84 to encode information - the $H$-, $V$-, $D$- and $A$-polarised states of a single photon \cite{Khokhlov2016BB84}. He claims this is through weak measurement - where weak coupling of a quantum variable to an ancilla allows data about a quantum state to be obtained without collapsing the state. However, weak measurement typically only obtains a small amount of information per measurement, so a large number of identically-prepared quantum objects are needed to obtain this fully. An alternative proposal given by Aharonov et al discusses the possibility of performing a weak measurement on a single particle \cite{Aharonov2014Foundations} - however, this is still subject to Busch's limit on information gained for a given disturbance \cite{Busch2009Limit}.

\begin{figure}
    \centering
    \includegraphics[width=\linewidth]{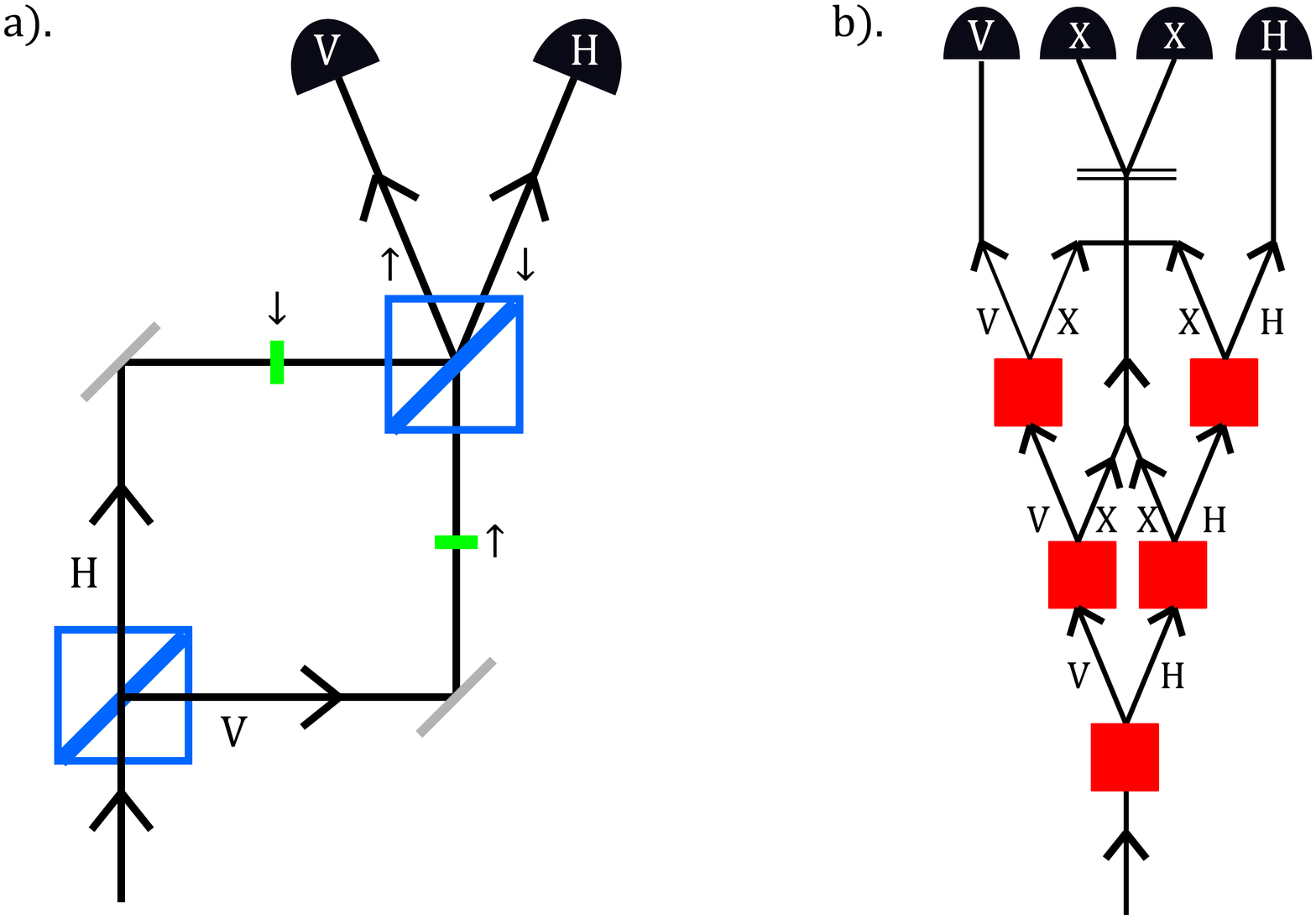}
    \caption{Diagrams of a) the single-interferometer device, which forms the building blocks of the attack protocol, and b) the layout formed of these, where each red square is an interferometer of the sort given in a. As can be seen, once initially split by the first interferometer, the $H$ and $V$ components remain these separate components, and so never travel the paths marked X. Therefore, only two of the four detectors ever receive photons, making the protocol useless for determining the full initial state for a single photon.}
    \label{fig:Figs}
\end{figure}

Khoklhov previously gave an interferometric device that he claims allows a weak measurement to tell the path a photon travelled via, without disturbing the state of that photon (see Fig.\ref{fig:Figs}a) \cite{Khokhlov2016Scheme}. For this, a photon of state $\alpha\ket{H}+\beta\ket{V}$ goes through a polarising beamsplitter (PBS), which transmits $H$- and reflects $V$-polarised components. The two components each travel down a respective arm, where they have a momentum-kick applied to them, such that their eventual arrival position on the second PBS isn't affected - here, the $H$-component gets a downwards kick, and the $V$-component an upwards kick. The two components meet at the second PBS, recombine, and then exit, but the difference in momentum allows, at a far distance from the second PBS, the respective components to be identified. Khoklhov claims, by using many photons, this allows the probability amplitudes of the two polarisation components to be determined, and so the preparation state (in truth, we only get the moduli-squared of those components, and so the classical balance of probabilities). Further, Khoklhov implies that the photon emitted into one of the two distinguishable far-field paths would still be in its original state, rather than collapsed to either $H$ or $V$. This is also incorrect - as the polarisation becomes entangled with the momentum degree-of-freedom, the collapse to either upwards- or downwards-momentum causes the simultaneous collapse to either $H$- or $V$-polarisation.

In this paper, Khoklov then makes the bold claim that, using these single-interferometer units as building blocks, he can make a device which can perfectly distinguish between the four basis states used in BB84 - $H$, $V$, $D$ (or $\frac{H+V}{\sqrt{2}}$) and $A$ (or $\frac{H-V}{\sqrt{2}}$). Given the security of BB84 rests on the quantum assumption that, even with an optimal choice of measuring basis, one cannot distinguish between all four bases perfectly (thus bounding an eavesdropper's potential knowledge), this claim threatens the security of one of the most well-known QKD protocols. 

Khoklhov claims this is possible by taking the separated outputs of one of his single-interferometer devices, then putting each output through another device. Those outputs which disagree with their original polarisation-determination (due to his assumption that the photon exits in its original state) are then combined - the components which agree with their original polarisation are put through another device, and again have any further outputs which disagree combined. The combined beam is then put through a polarising beamsplitter in the $A-D$ basis, with its two outputs sent to detectors, alongside the other two outputs (that always agreed with their initially-determined polarisation). We present this in Fig.\ref{fig:Figs}b. Khoklhov claims that if the input photon's initial state is $D$ ($A$) it will end up in the inner left (right) detector. This claim is false.

Let us examine the path of the input photon using standard quantum mechanics (as represented by Bra-Ket notation). We can describe the action of one of Khoklhov's single-interferometer devices (pre-detection), as given in Fig.\ref{fig:Figs}a, on a single polarisation-encoded photon qubit by
\begin{equation}
    \alpha\ket{H}+\beta\ket{V}\rightarrow\alpha\ket{H,\downarrow}+\beta\ket{V,\uparrow}
\end{equation}

This effectively entangles the polarisation and path degrees of freedom. Applying this to the larger set-up (as given in Fig.\ref{fig:Figs}b), we then see the two components ($H$ and $V$) then act as separate for the remainder of the chain, obeying

\begin{equation}
    \alpha\ket{H}+\beta\ket{V}\rightarrow\alpha\ket{H,R,R,R}+\beta\ket{V,L,L,L}
\end{equation}
where $L$ and $R$ describe the paths on the figure after each interferometer. This means the inner paths, $\ket{R,L}$, $\ket{L,R}$, $\ket{L,L,R}$ and $\ket{R,R,L}$ are never explored, so the two inner detectors never click. Given the determination of the single photon's polarisation qubit in the attack is predicated on these two detectors being able to click (being taken to represent $\frac{H+V}{\sqrt{2}}$ and $\frac{H-V}{\sqrt{2}}$), this shows the attack does not work - as expected, standard quantum mechanics preserves the security of BB84 from this attack.

We finally give a more proper account of the result of using true weak measurements to attempt Khoklhov's scheme, using the description of weak measurement from \cite{tamir2013introduction}.

To do this, in each of Khoklhov's apparatuses, we couple our photon's polarisation

\begin{equation}
\begin{split}
   \ket{\psi}&\in\{\ket{H};\ket{V};\frac{\ket{H}+\ket{V}}{\sqrt{2}};\frac{\ket{H}-\ket{V}}{\sqrt{2}}\}\\
   &=\alpha\ket{H}+\beta\ket{V}
   \end{split}
\end{equation}

with the pointer (the photon's momentum)

\begin{equation}
    \ket{\phi}=\ket{\phi_d}=\int_p \phi(p)\ket{p}dp
\end{equation}

where $p$ is the vertical momentum of the photon. $\hat{P}_d$ is the momentum operator such that $\hat{P}_d\ket{p}=p\ket{p}$.

We assume $\phi(p)$ has a Gaussian distribution around 0 (input vertical momentum), such that
\begin{equation}
    \phi(p)=e^{-p^2/4\sigma^2}/\sqrt{2\pi\sigma^2}
\end{equation}

If we define the polarisation-distinguishing operator
\begin{equation}
    \hat{A}=\ketbra{H}{H}-\ketbra{V}{V}
\end{equation}
we can consider an interaction Hamiltonian between the two
\begin{equation}
    \hat{H}_{int}=g(t)\hat{A}\otimes\hat{X}_d
\end{equation}
where $\hat{X}_d$ is the operator conjugate to $\hat{P}_d$ such that $[\hat{P}_d,\hat{X}_d]=i\hbar$, and $g(t)$ is the coupling function such that
\begin{equation}
    \int^T_0g(t)dt=1
\end{equation}
for coupling time $T$.

This means, applying this Hamiltonian
\begin{equation}
    e^{i\hat{H}t/\hbar}\ket{\psi}\otimes\ket{\phi}
\end{equation}
we see for each of $\ket{H}\otimes\ket{\phi(p)}$, $\ket{V}\otimes\ket{\phi(p)}$, the Hamiltonian takes $\hat{P}_d$ to $\hat{P}_d+1$, $\hat{P}_d-1$ respectively, as
\begin{equation}
    \begin{split}
        \hat{P}_d(T)-\hat{P}_d(0)=\int_0^T\frac{i}{\hbar}[\hat{H},\hat{P}_d]dt\in\{+1,-1\}
    \end{split}
\end{equation}

Therefore, the corresponding transformation is
\begin{equation}
\begin{split}
    &e^{i\hat{H}t/\hbar}\ket{\psi}\otimes\ket{\phi(p)}\\
    &=\alpha\ket{H}\otimes\ket{\phi(p-1)}+\beta\ket{V}\otimes\ket{\phi(p+1)}\\
    &=\int_p\big(\alpha\ket{H}\otimes\phi(p-1)+\beta\ket{V}\otimes\phi(p+1)\big)\ket{p}dp
\end{split}
\end{equation}

The above wavefunctions $\phi(p-1)$ and $\phi(p+1)$ need to overlap each other for the measurement to be weak - and so need to have high variance, $\sigma$. The higher the variance, the weaker the measurement - if these Gaussian wavefunctions don't overlap, them the measurement is strong. Given $\sigma$ is initially defined from the vertical momentum of the photon, this means the photon input into the system must also have high $\sigma$.

The effect of the strength of the measurement can be most readily seen when observe and collapse the pointer to a specific momentum-value, $p_0$, to read out our weak measurement, which gives
\begin{equation}
\Big(e^{-{\frac{(p_0-1)^2}{4\sigma^2}}}\alpha\ket{H}+e^{-{\frac{(p_0+1)^2}{4\sigma^2}}}\beta\ket{V}\Big)\otimes\ket{p_0}    
\end{equation}
where both the coefficients on $\ket{H}$ and $\ket{V}$ are biased slightly depending on where they are in relation to $p_0$. This makes sense, analogous to how measuring an eigenvalue for an observable collapses the measured state to the relevant eigenstate - the only difference here is the variance $\sigma$ providing some uncertainty in that measurement.

The far-field vertical position of the photon will depend on the vertical momentum of the photon, as Khoklhov rightly says. However, the variance in this momentum (which must be large enough to allow overlap between the momenta for up and for down in order for the measurement to be weak, and the final polarisation state to not have changed too far) means that these positions must overlap heavily too. While, in the limit of many identically-prepared photons, we could obtain information about whether the polarisation-state was $\ket{H}$, $\ket{V}$, or a superposition of the two, we cannot gain this for a single run without inducing collapse. Therefore, a protocol built up of several of these devices, as Khoklhov's attack protocol is, either doesn't work due to collapse (as we show with the standard quantum-mechanical approach above), or gains effectively no information about the polarisation state of the photon, making it useless as an attack.

\textit{Acknowledgements---}
This work was supported by the Engineering and Physical Sciences Research Council (Grants EP/P510269/1, EP/T001011/1, EP/R513386/1, EP/M013472/1 and EP/L024020/1).

\bibliographystyle{unsrt}
\bibliography{ref.bib}
\end{document}